\documentclass{article}
\pdfoutput=1

\usepackage[utf8x]{inputenc}
\usepackage[T1]{fontenc}
\usepackage{dsfont}
\usepackage{amstext}
\usepackage{amsfonts}
\usepackage{amsmath}
\usepackage{graphicx}
\usepackage[extendedchars]{grffile}
\usepackage{setspace}
\usepackage{gensymb}
\usepackage{color}
\usepackage[bookmarks=false]{hyperref}
\usepackage{doi}
\usepackage[numbers]{natbib}
\usepackage[affil-it]{authblk}

\PrerenderUnicode{ż}\PrerenderUnicode{ó}\PrerenderUnicode{ł}\PrerenderUnicode{ć}\PrerenderUnicode{ę}\PrerenderUnicode{ś}\PrerenderUnicode{ą}\PrerenderUnicode{ź}\PrerenderUnicode{ń}\PrerenderUnicode{Ż}\PrerenderUnicode{Ó}\PrerenderUnicode{Ł}\PrerenderUnicode{Ć}\PrerenderUnicode{Ę}\PrerenderUnicode{Ś}\PrerenderUnicode{Ą}\PrerenderUnicode{Ź}\PrerenderUnicode{Ń}

\newcommand{\abs}[1]{\left|#1\right|}

\newcommand{\dd}{\mathrm d}

\newcommand{\FFF}{\mathcal F}

\title{
Gluon distributions from Oliveira-Martin-Ryskin combined BFKL+DGLAP evolution equations.}

\author{Dawid Toton\thanks{\texttt{dawid.toton@ifj.edu.pl}}}

\affil{Instytut Fizyki Jądrowej im. H. Niewodniczańskiego \\
Radzikowskiego 152, 31-342 Kraków, Poland
}

\begin{document}
\maketitle

\begin{abstract}
Kwiecinski, Martin, Stasto \cite{Kwiecinski:1997ee} argue for inclusion of DGLAP terms into BFKL evolution of unintegrated gluon density.
The equation was reformulated by Oliveira, Martin, Ryskin \cite{deOliveira:2014cua}
employing the opening angle $\theta = \frac {k}{xp}$ as the evolution variable.
It leads to a description of a $\theta$-integrated gluon density.
This paper is a numerical study of these two similar combined BFKL+DGLAP formulations.
It is a demonstration of feasibility of the new approach.
The different ways of subtracting the contribution common for BFKL and DGLAP
proposed in  \cite{Kwiecinski:1997ee} and  \cite{deOliveira:2014cua}
are compared.
The numerical tests confirm that the $\theta$ variable is a more natural evolution variable
for this kind of equation.
\end{abstract}

\section{Introduction}

The framework of collinear factorisation
and its DGLAP description of parton distributions proved overly successful where applicable.
On the other hand, in the regime where $k_t$-factorisation is the proper tool,
we look for the effects of low-$x$ physics and there are some hints of its dynamics seen\cite{
vanHameren:2014ala,Dusling:2012cg,Ducloue:2013bva}.

One of the advancements in the field was realisation that the original BFKL formulation
\cite{Kuraev:1977fs, Balitsky:1978ic}
included regions of phase space which violate the assumed kinematics of the process.
Thus it was understood that the kernel should include
the so-called kinematical consistency requirement, which incorporates corrections of all orders \cite{Andersson:1995ju,KMS96}.
It is an important change, since it significantly reduces growth of the gluon densities \cite{Andersson:1995ju,KMS96}.

Among many attempts to enrich the BFKL picture with soft emissions
was development of the CCFM evolution equation \cite{Catani:1989sg, Catani1990339}.
It is an interpolation between BFKL and DGLAP achieved in presence of colour coherence.
A much simpler proposal of Kwiecinski, Martin and Stasto \cite{Kwiecinski:1997ee}
was to use a hybrid BFKL+DGLAP evolution kernel,
 specially crafted to avoid double counting.

A recent paper by Oliveira, Martin and Ryskin \cite{deOliveira:2014cua}
continues with this scheme and brings a new insight.
The crucial observation is that
the inclusion of the kinematical consistency constraint,
makes the set of points entering the evolution kernel more convex, so that
now we have a freedom to alter the direction of evolution of the gluon distribution.
The authors have chosen the angle $\theta = \frac k {xp}$ as a new evolution variable.
This made a particularly compelling formulation of the combined BFKL+DGLAP kernel.
Moreover, a two changes with respect to the original formulation of Kwiecinski et al. were included:
(1) the double-counting removal term was broadened to match the kinematical constraint applied to BFKL kernel,
(2) an extra term was proposed to help the BFKL kernel with energy-momentum conservation.

We adapt these hybrid evolution equations for numerical calculations
and study the behaviour of the implementations to evaluate usefulness of the proposed variable change.

\section{Combined BFKL+DGLAP evolution}

The combined BFKL+DGLAP evolution of \cite{Kwiecinski:1997ee} with quark contribution neglected
reads\footnote{
See also Eq. (10) of \cite{Kwiecinski:1997ee} and Eq. (3.1) of \cite{Kutak:2012rf}.
}:

\begin{equation}\label{bfkl+dglap}
\begin{split}
\FFF(x, k^2) &=
\FFF_0(x, k^2)
  \\&+
\bar\alpha_s
\int_{x}^1 \frac {\dd z}{z}
\int_{k_0^2}^\infty \frac{\dd l^2}{l^2}
\left(
\frac
  {l^2\FFF(\frac x z, l^2)\Theta(k^2 - z l^2) - k^2 \FFF(\frac x z , k^2)}
  {\abs{l^2 - k^2}}
+\frac
  {k^2 \FFF(\frac x z , k^2)}
  {\sqrt{4 l^4 + k^4}}
\right)
 \\&+
\frac{\bar\alpha_s}{k^2}
\int_{x}^1 \frac {\dd z}{z}
\left(
\frac{z P_{gg}(z)}{2 N_c} -1
\right)
\int_{k_0^2}^{k^2} \dd l^2 \FFF\left(\frac x z, l^2\right)
\end{split}
\end{equation}

With the leading-order gluon-gluon splitting function \cite{ellis2003qcd}
\begin{equation}\label{Pgg}
P_{gg}(z) =
2 C_A
\left(
\frac {z}{(1-z)_+} + \frac {1-z}z + z(1-z)
\right)
+\delta(1-z)
\frac {11 C_A - 4 n_f T_R}
{6}
\end{equation}
inserted, the DGLAP term acquires the following explicit form
\begin{equation}
\begin{split}
&\int_{x}^1 \frac {\dd z}{z}
\left(
\frac{z P_{gg}(z)}{2 N_c} -1
\right)
\int_{k_0^2}^{k^2} \dd l^2 \FFF(\frac x z, l^2)
=\\&
\int_{x}^1 \dd z
\int_{k_0^2}^{k^2} \dd l^2
\Biggl(
  \left(
  \frac{C_A}{N_c}
  \left(
  \frac {z}{(1-z)} + \frac {1-z}z + z(1-z)
  \right)
  -
    \frac {1}{z}
  \right)
  \FFF(\frac x z, l^2)
-
  \frac {C_A}{N_c}
  \frac {1}{(1-z)}
   \FFF(x, l^2)
\Biggr)
+ \\&
  \left(
    \frac {11 C_A - 4 n_f T_R}
    {12 N_c}
  +
  \frac {C_A}{N_c}
  \log\left(1-x\right)
  \right)
  \int_{k_0^2}^{k^2} \dd l^2 \FFF(x, l^2)
\end{split}
\end{equation}

Note the term with $\log(1-x)$, which is a piece of the normalisation term originating from the plus prescription.

\section{Evolution in angle}

The distribution $\FFF(x, k^2)$ used in Eq. \ref{bfkl+dglap}, when rescaled,
becomes $f(x, k) = k^2 \FFF(x, k^2)$, which is the gluon density used in \cite{deOliveira:2014cua}.
In the paper, Oliveira, Martin and Ryskin developed
an equation of gluon evolution with angle\footnote{
$k_t$ and $k_t'$ of \cite{deOliveira:2014cua} are denoted respectively $k$ and $l$ here.
} $\theta= \frac {k}{xp}$,
where $p$ is the momentum of the incoming hadron.
It describes the integrated distribution
$x \hat g(x, \theta) = \int^{\theta^2}\frac {\dd {\theta'}^2}{{\theta'}^2} \hat f(x, \theta')$.
This new angle-dependent gluon density is defined by
\begin{equation}
\hat f(x, \theta) = f(x, x p \theta)
\end{equation}

To simplify accuracy control in the numerical procedure, we avoid taking derivatives.
Hence, remembering that
\begin{equation}
\frac {x \partial \hat g(x,\theta)}{\partial \log \theta^2} = \hat f(x, \theta)
\end{equation}
we keep the equation in the integral from\footnote{
The $\int\dd^2{k_t'}$ integral in Eq. (1) of \cite{deOliveira:2014cua} brings an extra $\pi$ factor,
which makes it inconsistent with BFKL, so it is omitted here.
Also, note that in our notation the $\Theta(z - x)$ limit is not included in the splitting $P(z)$.
}
\begin{equation}\label{Ryskin}
\begin{split}
\hat f(x, \theta) &= \hat f_0(x, \theta) + \\&
\frac {\alpha_s}{2\pi}\biggl(
\int_0^\infty \dd {l}^2
\int_{\theta_{min}(x, l)}^{\theta_{max}(x, l)} \frac{\dd \theta'}{\theta'}
\bar K(k, l) f(x', l)
-
\int_0^1\dd z
\int_0^\infty \dd {l}^2 \bar K(k, l) f(x, l)
\\&
+
\int_x^1\dd z
P(z) \frac x z \hat g\left(\frac x z, z \theta\right)
\biggr)
\end{split}
\end{equation}

The $\dd \theta'$ integration limits $\theta_{min}(x, l)$ and $\theta_{max}(x, l)$ require a comment.
This integral originates from $\int_x^1 \dd z$
and it must retain the property that $z<1$
for the new BFKL kernel to be correct.
The authors of \cite{deOliveira:2014cua} decided to set $\theta_{max}(x, l) = \theta$,
which unfortunately admits $z\theta>\theta$ in the integrand,
since $z = x \frac p l \theta' = \frac {k\theta'} {l\theta} $.
On the other hand, the original BFKL kernel integration area is reproduced if we take
\begin{equation}
\theta_{min}(x, l) = \frac l{p}
\quad\quad
\theta_{max}(x, l) = \frac l{x p}
\end{equation}
instead. As will be made clear below, this correction does not prevent $\theta$ to be
employed as an evolution variable along the lines of the discussion in \cite{deOliveira:2014cua}.
The lower integration limit proposed in \cite{deOliveira:2014cua},
$\theta_{min}(x, l) = \theta_0(x) = \frac {k_0}{xp}$
seems an appealing choice,
since this $l$-independent cut makes the evolution equation more elegant.
Sadly, this would cause sizeable contributions from the
 low-$k$, high-$x$ region to be neglected,
 so the idea has to be abandoned.

Since the kinematical consistency constraint limits ${l}^2$ by $\frac {k^2} z$,
we can safely impose a maximal momentum $k_{max}$ to approximate the $\dd {l}^2$ integrals.
Moreover, we are currently not interested in modelling the infra-red region.
Thus, instead of the smooth extrapolation for $f(x, k<k_0)$ proposed in \cite{Askew:1993jk, deOliveira:2014cua}, we introduce a simple cutoff $k_0$.
As a consequence,
\begin{equation}
\hat f(x, \theta) = 0 \quad \text{for} \quad \theta < \frac {k_0}{x p} = \theta_0(x)
\end{equation}
Therefore, we can also have a lower limit in the integral needed for $\hat g$, which becomes
\begin{equation}
x \hat g(x, \theta) = \int_{\theta_0^2(x)}^{\theta^2}\frac {\dd {\theta'}^2}{{\theta'}^2} \hat f(x, \theta')
\end{equation}

With all the integration limits in place we have
\begin{equation}
\begin{split}
\hat f(x, \theta) &= \hat f_0(x, \theta) + \\&
\bar\alpha_s
\biggl(
\int_{l / p}^{l/(x p)} \frac{\dd \theta'}{\theta'}
\int_{k_0^2}^{k_{max}^2} \dd l^2
\frac 1 {2N_c}
\bar K(k, l) f(x', l)
-
\int_0^1\dd z
\int_{k_0^2}^{k_{max}^2} \dd l^2
\frac 1 {2N_c}
\bar K(k, l) f(x, l)
\\&
+
\int_x^1\frac{\dd z}z
\frac {z P(z)} {2N_c}
\int_{z^2\theta_0^2(x)}^{(z\theta)^2} \frac{\dd {\theta'}^2}{{\theta'}^2}
\hat f\left(\frac x z, \theta'\right)
\biggr)
\end{split}
\end{equation}

Following Eq. \ref{Pgg} we expand the splitting function
(with quarks neglected),
and obtain a fully explicit form of the evolution equation
\begin{equation}
\hat f(x, \theta) = \hat f_0(x, \theta) +
\bar\alpha_s \left(I_{BFKL-dc} + I_{ec} + I_{DGLAP} + I_{norm}\right)
\end{equation}

The four integrals, with the help of the notation $z = x \frac p l \theta'$ and $k= x p \theta$, are
(a) BFKL with kinematical consistency constraints with doubly counted double-logarithmic part removed
\begin{equation}\label{BFKL-dc}
\begin{split}
&I_{BFKL-dc} =
\int_{k_0^2}^{k_{max}^2}\frac {\dd l^2} {l^2}
\int_{l / p}^{l/(x p)} \frac{\dd \theta'}{\theta'}
\\&
k^2
\left(
\frac{\Theta(k^2 - z l^2)\hat f(\frac l {p\theta'}, \theta') -
  \hat f\left(\frac l {p\theta'}, z \theta\right)}{\abs{l^2 - k^2}}
+
\frac{\hat f\left(\frac l {p\theta'}, z \theta \right)}{\sqrt{4 l^4 + k^4}}
-
\frac{\Theta(k^2 - z l^2) \hat f(\frac l {p\theta'},\theta')}{k^2}
\right)
\end{split}
\end{equation}

(b) a term subtracted to restore energy-momentum conservation of BFKL
\begin{equation}\label{Iec}
\begin{split}
&I_{ec} =
-
\int_{0}^1\dd z
\int_{k_0^2}^{k_{max}^2} \frac {\dd l^2} {l^2}
\\&
k^2
\left(
\frac{\Theta(k^2 - z l^2)\hat f_\theta\left(x, \frac {l} {x p}\right) - \hat f(x, \theta)}{\abs{l^2 - k^2}}
+
\frac{\hat f(x, \theta)}{\sqrt{4 l^4 + k^4}}
-
\frac{\Theta(k^2 - z l^2) \hat f_\theta\left(x, \frac {l} {x p}\right)}{k^2}
\right)
\end{split}
\end{equation}

(c) the $z$-dependent part of the DGLAP evolution kernel
\begin{equation}
\begin{split}
&I_{DGLAP} =
  \frac {C_A}{N_c}
  \int_{x}^1 \dd z
  \int_{z^2\theta_0^2(x)}^{\theta^2} \frac{\dd {\theta'}^2}{{\theta'}^2}
\\&
  \Biggl(
    \left(
    \frac {z}{(1-z)} + \frac {1-z}z + z(1-z)
    \right)
    \Theta(z^2 \theta^2 - {\theta'}^2)
    \hat f\left(\frac x z, \theta'\right)
  -
    \frac {\Theta({\theta'}^2 - \theta_0^2(x))}{(1-z)}
    \hat f\left(x, \theta'\right)
  \Biggr)
\end{split}
\end{equation}

(d) the DGLAP normalisation from the $\frac 1 {1-z}$ pole together with the integrated $\delta(1-z)$ term
\begin{equation}
\begin{split}
&I_{norm} =
    \left(
      \frac {11 C_A - 4 n_f T_R}
      {12 N_c}
    +
    \frac {C_A}{N_c}
    \log\left(1 - x\right)
    \right)
    \int_{\theta_0^2(x)}^{\theta^2} \frac{\dd {\theta'}^2}{{\theta'}^2}
    \hat f\left(x, \theta'\right)
\end{split}
\end{equation}

\begin{figure}\label{evolution-point}
\centerline{
\includegraphics[width=8cm]{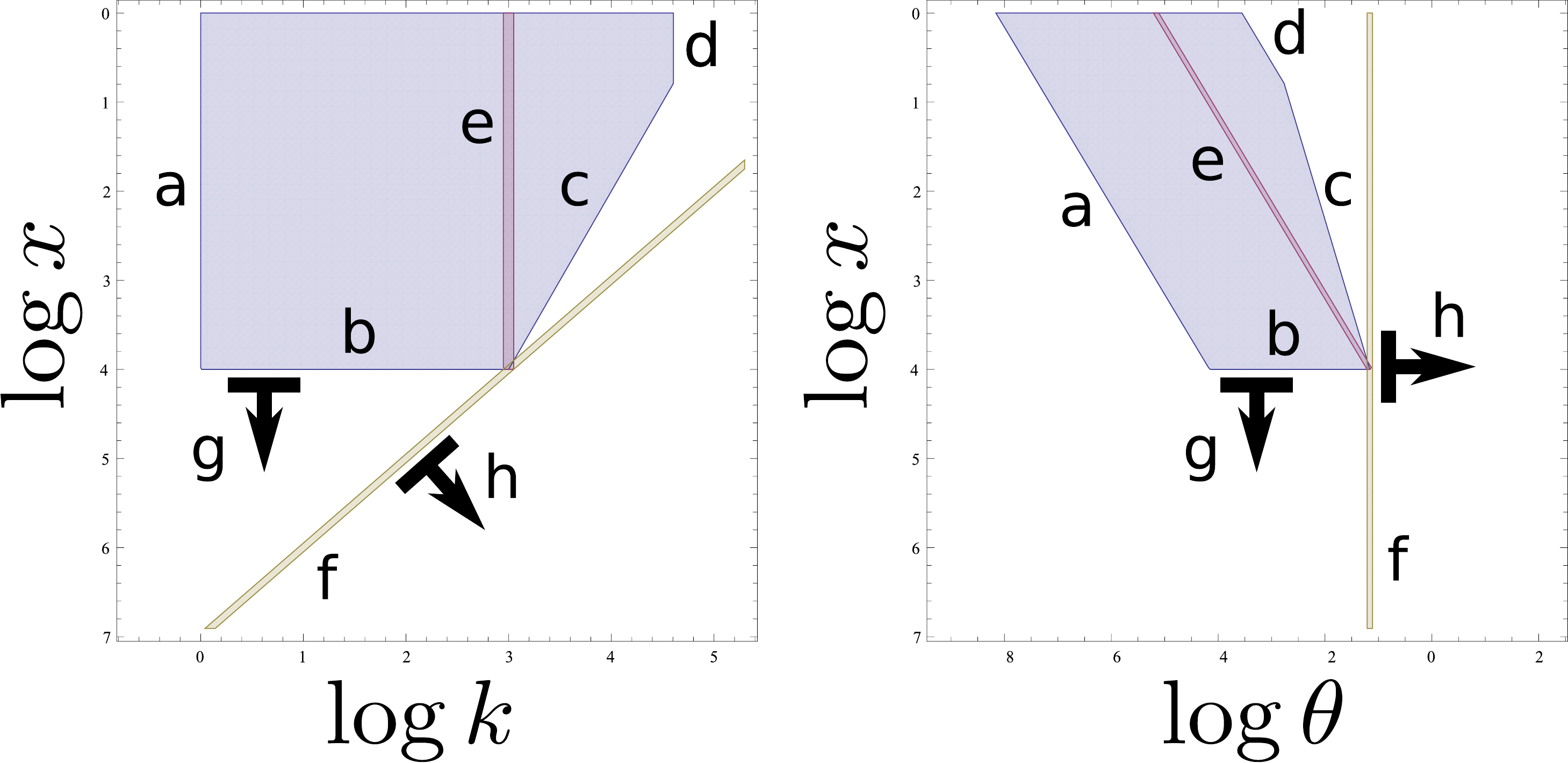}
}
\caption{Depiction of action of the kernel for a particular $(x,k)$ point
on the $\log x, \log k$ plane (left) and
$\log x, \log \theta$ plane (right).
The highlighted lines have the following meaning:
(a) infrared cutoff,
(b) a line of constant $x$, which contributes to the virtual DGLAP terms,
(c) a cut imposed by kinematical consistency requirement,
(d) the $k_{max}$ cutoff applied for numerical convenience,
(e) (pink line) the points contributing to the virtual terms of the BFKL kernel,
(f) a line of constant $\theta$.
Arrows:
(g) the usual evolution with $x$,
(h) the new evolution directed by $\theta$;
blue area: the points contributing to real emissions of BFKL.
 }
\end{figure}

The function $\hat f_\theta(x, \theta_1)$ put in the $I_{ec}$ term (Eq. \ref{Iec}) is
exactly the same as $\hat f(x, \theta_1)$ for $\theta_1 < \theta$ and
for greater values of $\theta_1$ it becomes some approximate guess of $\hat f$ instead.
The authors of \cite{deOliveira:2014cua} suggest that the $\hat f_\theta$ function
could be taken from a previous, less accurate solution and improved in a self-consistent way.
In our calculations, which perform truly $\theta$-directed evolution,
we either omit $I_{ec}$ or insert an extra $\Theta\left(\theta - \frac l {xp}\right)$ factor
 into the integrand in $I_{ec}$.
This treatment weakens the correction compared to the original one,
but is a necessary change to prevent reaching for $\hat f_\theta$
where the values of
$\hat f\left(x, \frac l {xp} > \theta\right)$ are unavailable.

Let us see which values of $\FFF(x, k)$ and $\hat f(x, \theta)$
become available as the evolution progresses.
Consider the $(x, k)$ and $(x, \theta)$ planes,
shown on Fig. \ref{evolution-point}.
Upper-left corners of both plots correspond to low $x$ and low $k$,
where the evolution starts.
The equations for the gluon densities enjoy
the property that the domain can be ordered
so that
to determine the right value
at some point it suffices to
have the solution at the points considered previous.
These points are depicted as the blue areas on Fig. \ref{evolution-point}.
The point for which we perform a calculation
is where the lines \emph{b}, \emph{c}, \emph{e}, \emph{f} meet.
The evolution kernel is an integral over the blue area.
In the BFKL approximation
the
hard emissions, $z\approx 0$,
dominate the integral.
This is the upper edge of the plot.
A natural numerical procedure which follows the evolution
extends the current solution across the line \emph{b} very easily.
This is because it depends on the already well established solution around $x\approx 1$.
On the other hand,
when one wants to take account of soft gluon emissions, $z\approx 1$,
the situation changes.
This pole of the splitting function
makes the area in the vicinity of line \emph{b}
a significant contribution.
Now, the whole solution along the line $b$ needs to be self-consistent.
It thus becomes inconvenient to determine the right value for all the $k$ points simultaneously
and the numerical treatment becomes more intricate.
In this region of the phase space
the gluons build up their transversal momentum.
This is closely related to the fact that the momentum transfer $Q^2$ is the evolution variable of DGLAP.

Now, consider the same integration area on the $(x, \theta)$ plane as proposed in \cite{deOliveira:2014cua}.
To the right of the line \emph{c} are the points with $l^2 > \frac {k^2} z$.
Not satisfying the kinematical consistency, they are excluded
giving us the freedom to change the direction of the evolution.
The choice of $\theta = \frac k {xp}$, marked with arrow \emph{h} on Fig. \ref{evolution-point},
greatly reduces the above-mentioned numerical complexity.
In this new scheme, only the proximity of the calculated point
needs careful treatment, the rest of the $\theta = const$ line remains disentangled.
We will discuss this point further below.

\section{Solutions and their properties}

The equations \ref{bfkl+dglap} and \ref{Ryskin} were implemented
to solve them by iterative refinements.
The algorithm is similar to the one described in \cite{Kutak:2013kfa};
it employs a standard Monte-Carlo integration procedure and bilinear interpolation
of the solution.
Eq. \ref{bfkl+dglap} is solved evolving the gluon density along $x$ while
 with Eq. \ref{Ryskin} both $x$ and $\theta$ directions are studied.
We have found that the weakened form of the $I_{ec}$ correction we employed,
brings a change of the solution that is small compared to the accuracy of our calculations.
Thus,
apart from the already mentioned minor modifications of the original formulation of Oliveira et al,
we also exclude the $I_{ec}$ term.
As the recipes for double-counting removal
proposed in \cite{Kwiecinski:1997ee} and \cite{deOliveira:2014cua} differ,
we modify the last term in $I_{BFKL-dc}$ (Eq. \ref{BFKL-dc})
and observe that the solution consequently converges to the result of Eq. \ref{bfkl+dglap}.

\begin{figure}\label{results-bfkl+dglap-comparison}
\centerline{
\includegraphics[trim=1cm 0 3cm 0,width=9cm]{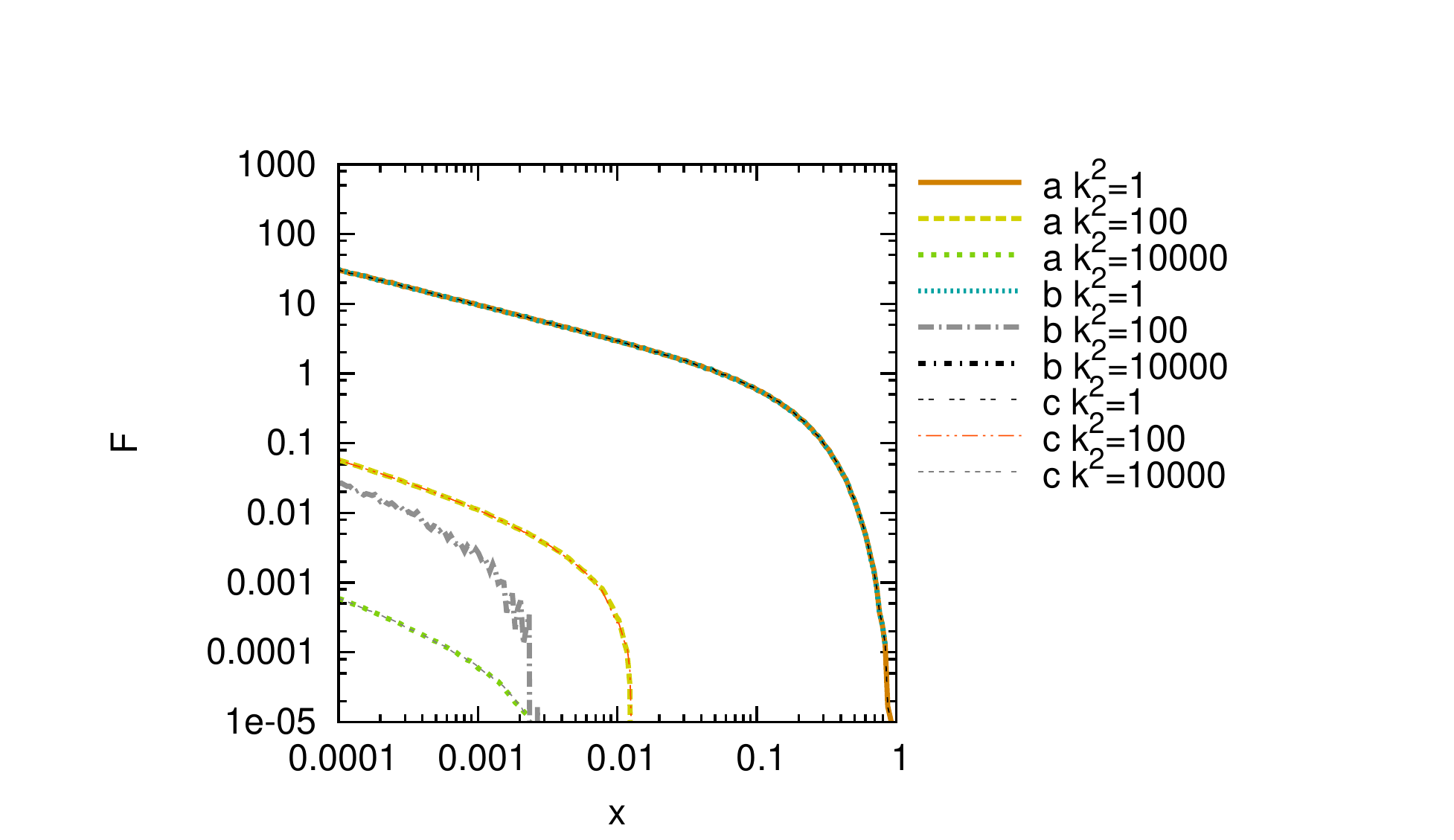}
\includegraphics[trim=1cm 0 3cm 0,width=9cm]{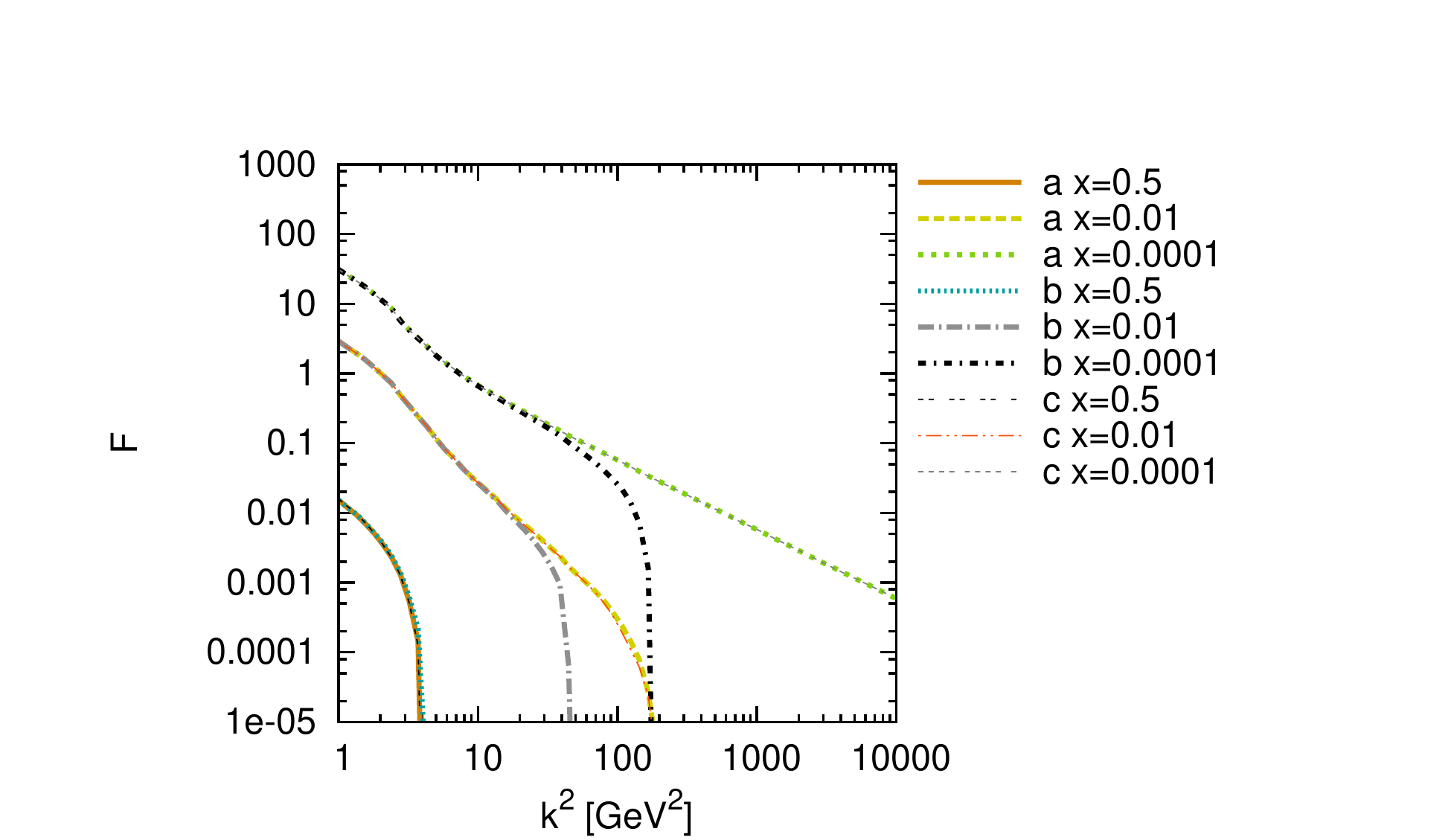}
}
\caption{Gluon distributions $\FFF(x, k^2)$ resulting from: Eq. \ref{Ryskin}
with the $I_{dc}^{DGLAP}$ modification (\emph{a} and \emph{c});
Eq. \ref{bfkl+dglap} (\emph{b} and \emph{c}).
Lines \emph{c} show a correctly converged result;
\emph{a} and \emph{b} are obtained with the algorithm limited to 8 refinements
to compare the behaviour of the solvers.
}
\end{figure}

The implemented solvers deliver results of Eq. \ref{Ryskin} that are as good as
the ones obtained from Eq. \ref{bfkl+dglap}.
They are plotted on Fig. \ref{results-bfkl+dglap-comparison} with lines \emph{c}.
More precisely, these lines correspond to
the density $\FFF$ from
both
the solution of Eq. \ref{bfkl+dglap}
and the one of Eq. \ref{Ryskin} with a modified $I_{dc}$ term (Eq. \ref{I_dc_DGLAP}, discussed below).
If they were put on Fig. \ref{results-bfkl+dglap-comparison} separately,
the lines would be nearly indistinguishable.
This reassures us that the two significantly different solvers used have attained acceptable accuracy.

The above-mentioned modification of $I_{dc}$ that makes the two equations equivalent
is only a change made to the theta function.
Namely,
the recipe found in \cite{deOliveira:2014cua}
\begin{equation}\label{I_dc_BFKL}
I_{dc}^{BFKL} =
\int_{k_0^2}^{k_{max}^2}\frac {\dd l^2} {l^2}
\int_{l / p}^{l/(x p)} \frac{\dd \theta'}{\theta'}
\Theta(k^2 - z l^2) \hat f(\frac l {p\theta'},\theta')
\end{equation}
can have the range of $l$ tightened
\begin{equation}\label{I_dc_DGLAP}
I_{dc}^{DGLAP} =
\int_{k_0^2}^{k_{max}^2}\frac {\dd l^2} {l^2}
\int_{l / p}^{l/(x p)} \frac{\dd \theta'}{\theta'}
\Theta(k^2 - l^2) \hat f(\frac l {p\theta'},\theta')
\end{equation}
to match the doubly logarithmic contribution of DGLAP, as subtracted in \cite{Kwiecinski:1997ee}.
The first option, $I_{dc}^{BFKL}$ admits $l\gg k$ for small $z$,
which brings a contribution which is beyond DGLAP and a bit problematic for BFKL too.
Namely, the real part of BFKL kernel in this regime 
 becomes
\begin{equation}
\int_{k_0^2}^{k_{max}^2}\frac {\dd l^2} {l^2}
\int_{l / p}^{l/(x p)} \frac{\dd \theta'}{\theta'}
\frac{k^2}{l^2}
\Theta(k^2 - z l^2) \hat f\left(\frac l {p\theta'},\theta'\right)
\end{equation}
and due to the $\frac{k^2}{l^2} \ll 1$ factor it becomes too small for $I_{dc}^{BFKL}$.
Looking at several numerical solutions,
we found that it is easy to encounter conditions for which
the term $I_{BFKL-dc}$ (Eq. \ref{BFKL-dc}), which follows Eq. \ref{I_dc_BFKL},
 becomes negative
and at some small $x$ the gluon density drops quickly.
Anyway, all the solutions seem to be well-behaved for the initial conditions used throughout this paper
\footnote{
The initial condition $f_0(x, \theta)$ in Eq. (8) of \cite{deOliveira:2014cua}
is
$\theta$-independent,
but we can allow here for other forms of $f_0$ at no cost.
}
\begin{equation}
\FFF_0(x, k) = x^{-B}(1-x)^5 \exp\left( - \left(\frac{k}{k_{ic}}\right)^2\right)
\end{equation}
and its equivalent
\begin{equation}
\hat f_0(x, \theta) =
  x^{-B}(1-x)^5
  \left(xp\theta\right) ^2 \exp\left( - \left(\frac{xp\theta}{k_{ic}}\right)^2\right)
\end{equation}
A choice of $k_{ic} = 1\text{ GeV}$ and $B=0.5$ we employed gives a reasonable gluon distribution.
The calculations were performed with the strong coupling kept at $\bar\alpha_s=0.2$
and $n_f = 4$ flavours.



As discussed above,
we expect that the ordering of the evolution along the $\theta$ axis
makes calculations easier.
To verify this, we limit our algorithms to 8 iterations
and compare the resulting inaccurate solutions.
These are plots \emph{a} (for $\theta$-directed evolution) and \emph{b} (for $x$)
on Fig. \ref{results-bfkl+dglap-comparison}.
At high $k$, the lines differ significantly.
With this small number of refinement steps,
the $\theta$-directed result is very close to the converged solution \emph{c},
while \emph{b} lacks high-$k$ gluons.
This is understandable,
since the evolution makes the gluons diffuse towards higher $k$.
As a consequence,
 the final refinements correct the density at the highest transverse momentum.

This last result demonstrates
that the idea presented in \cite{deOliveira:2014cua}
is not just a mere change of variables,
but is a good choice of a natural direction of the evolution.

\section{Summary}

We pursue the new route proposed by Oliveira et al and perform actual calculations of
the hybrid BFKL+DGLAP
gluon evolution in the angle $\theta$.
The differences between the original formulation and the new one are discussed.
We point out a likely need to improve the proposed way the doubly counted double-logarithmic contribution is subtracted.
The numerical calculations performed demonstrate that the new approach is feasible and worthy.

\section*{Acknowledgements}
The author wishes to thank Krzysztof Kutak
for pointing out the importance of the $x^{-B}$ factor in the initial distribution.
This work was supported by the NCBiR Grant No. LIDER/02/35/L-2/10/NCBiR/2011.

\bibliography{bfkldglap.bib}
\bibliographystyle{plainnat}


\end{document}